\documentclass{aastex62}
\hypersetup{linkcolor=red,citecolor=blue,filecolor=cyan,urlcolor=magenta}

\usepackage{xcolor}
\definecolor{dgray}{gray}{0.33}
\accepted{March 28, 2023}

\submitjournal{ApJL}

\shorttitle{W0336}
\shortauthors{Calissendorff et al.}
\begin{document}

\title{JWST/NIRCam discovery of the first Y+Y brown dwarf binary: WISE J033605.05$-$014350.4}

\correspondingauthor{Per Calissendorff}
\email{percal@umich.edu}

\author[0000-0002-5335-0616]{Per Calissendorff}
\affil{Department of Astronomy, University of Michigan, Ann Arbor, MI 48109, USA}
\author[0000-0003-1863-4960]{Matthew De Furio}
\affil{Department of Astronomy, University of Michigan, Ann Arbor, MI 48109, USA}
\author[0000-0003-1227-3084]{Michael Meyer}
\affil{Department of Astronomy, University of Michigan, Ann Arbor, MI 48109, USA}
\author[0000-0003-0475-9375]{Lo\"ic Albert}
\affiliation{D\'epartement de Physique and Observatoire du Mont-M\'egantic, Universit\'e de Montr\'eal, C.P. 6128, Succ. Centre-ville, Montr\'eal, H3C 3J7, Québec, Canada.}
\affiliation{Institut Trottier de Recherche sur les exoplan\`etes, Universit\'e de Montr\'eal}
\author[0000-0003-2094-9128]{Christian Aganze}
\affil{UC San Diego}
\author[0000-0002-6633-376X]{Mohamad Ali-Dib}
\affiliation{Institut Trottier de Recherche sur les exoplan\`etes, Universit\'e de Montr\'eal}
\affil{Center for Astro, Particle and Planetary Physics (CAP3),
New York University Abu Dhabi, UAE}
\author[0000-0001-8170-7072]{Daniella C. Bardalez Gagliuffi}
\affil{Department of Physics \& Astronomy, Amherst College, 25 East Drive, Amherst, MA 01003, USA}
\author{Frederique Baron}
\affiliation{D\'epartement de Physique and Observatoire du Mont-M\'egantic, Universit\'e de Montr\'eal, C.P. 6128, Succ. Centre-ville, Montr\'eal, H3C 3J7, Québec, Canada.}
\affiliation{Institut Trottier de Recherche sur les exoplan\`etes, Universit\'e de Montr\'eal}
\author[0000-0002-5627-5471]{Charles A. Beichman}
\affil{Jet Propulsion Laboratory}
\author[0000-0002-6523-9536]{Adam J. Burgasser}
\affil{University of California - San Diego}
%\author{Alban Ceau}
%\affil{Observatoire de la Cote d'Azur}
\author[0000-0001-7780-3352]{Michael C. Cushing}
\affil{Ritter Astrophysical Research Center, Department of Physics and Astronomy, University of Toledo, 2801 W. Bancroft Street, Toledo, OH 43606, USA}
\author[0000-0001-6251-0573]{Jacqueline Kelly Faherty}
\affil{Astrophysics Department, American Museum of Natural History, 79th Street at Central Park West, New York, NY 10024}
\author[0000-0002-2428-9932]{Cl\'emence Fontanive}
\affiliation{D\'epartement de Physique and Observatoire du Mont-M\'egantic, Universit\'e de Montr\'eal, C.P. 6128, Succ. Centre-ville, Montr\'eal, H3C 3J7, Québec, Canada.}
\affiliation{Institut Trottier de Recherche sur les exoplan\`etes, Universit\'e de Montr\'eal}
\author[0000-0001-5072-4574]{Christopher R. Gelino}
\affil{California Institute of Technology}
\author[0000-0002-8916-1972]{John E. Gizis}
\affil{University of Delaware}
\author[0000-0002-7162-8036]{Alexandra Z. Greenbaum}
\affiliation{IPAC, Mail Code 100-22, Caltech, 1200 E. California Blvd., Pasadena, CA 91125, USA}
\author[0000-0003-4269-260X]{J. Davy Kirkpatrick}
\affil{California Institute of Technology}
\author[0000-0002-3681-2989]{Sandy K. Leggett}
\affil{NOIRLab - Gemini North (HI)}
\author[0000-0003-1180-4138]{Frantz Martinache}
\affil{Université Côte d’Azur, Observatoire de la Côte d'Azur, CNRS, Laboratoire Lagrange, France}
\author{David Mary}
\affil{Université Côte d’Azur, Observatoire de la Côte d'Azur, CNRS, Laboratoire Lagrange, France}
\author[0000-0002-1721-3294]{Mamadou N'Diaye}
\affil{Université Côte d’Azur, Observatoire de la Côte d'Azur, CNRS, Laboratoire Lagrange, France}
\author[0000-0003-2595-9114]{Benjamin J. S. Pope}
\affiliation{School of Mathematics and Physics, The University of Queensland, St Lucia, QLD 4072, Australia}
\affiliation{Centre for Astrophysics, University of Southern Queensland, West Street, Toowoomba, QLD 4350, Australia}
\author[0000-0002-6730-5410]{Thomas L Roellig}
\affil{MS 245-6, NASA Ames Research Center, Moffett Field, CA 94035}
\author[0000-0001-9525-3673]{Johannes Sahlmann}
\affil{RHEA Group for the European Space Agency (ESA), European Space Astronomy Centre (ESAC), Camino Bajo del Castillo s/n, 28692 Villanueva de la Ca\~nada, Madrid, Spain}
\author[0000-0003-1251-4124]{Anand Sivaramakrishnan}
\affiliation{Space Telescope Science Institute, 3700 San Martin Drive, Baltimore, MD 21218, USA}
\affiliation{Astrophysics Department, American Museum of Natural History, 79th Street at Central Park West, New York, NY 10024}
\affiliation{Department of Physics and Astronomy, Johns Hopkins University, 3701 San Martin Drive, Baltimore, MD 21218, USA}
\author[0000-0002-5113-8558]{Daniel Peter Thorngren}
\affil{Universite de Montreal}
\author[0000-0001-7591-2731]{Marie Ygouf}
\affil{Jet Propulsion Laboratory}
\author[0000-0002-5922-8267]{Thomas Vandal}
\affiliation{D\'epartement de Physique and Observatoire du Mont-M\'egantic, Universit\'e de Montr\'eal, C.P. 6128, Succ. Centre-ville, Montr\'eal, H3C 3J7, Québec, Canada.}
\affiliation{Institut Trottier de Recherche sur les exoplan\`etes, Universit\'e de Montr\'eal}

\begin{abstract}
\noindent We report the discovery of the first brown dwarf binary system with a Y dwarf primary, WISE J033605.05$-$014350.4, observed with NIRCam on JWST with the F150W and F480M filters. We employed an empirical point spread function binary model to identify the companion, located at a projected separation of 0$\farcs$084, position angle of 295 degrees, and with contrast of 2.8 and 1.8 magnitudes in F150W and F480M, respectively. At a distance of 10\,pc based on its Spitzer parallax, and assuming a random inclination distribution, the physical separation is approximately 1\,au. Evolutionary models predict for that an age of 1-5 Gyr, the companion mass is about 4-12.5 Jupiter masses around the 7.5-20 Jupiter mass primary, corresponding to a companion-to-host mass fraction of $q=0.61\pm0.05$. Under the assumption of a Keplerian orbit the period for this extreme binary is in the range of 5-9 years. The system joins a small but growing sample of ultracool dwarf binaries with effective temperatures of a few hundreds of Kelvin. Brown dwarf binaries lie at the nexus of importance for understanding the formation mechanisms of these elusive objects, as they allow us to investigate whether the companions formed as stars or as planets in a disk around the primary.

\end{abstract}


\section{Introduction} \label{sec:intro}
\noindent Brown dwarfs are cool substellar objects with masses below the $\approx 80\,M_{\rm Jup}$ hydrogen burning limit \citep[e.g.][]{Burrows+2001}. Studies of the initial mass function for stars extend far into the substellar regime, even below 5~$M_{\rm Jup}$ \citep{Kirkpatrick+2019}, which sets a tight boundary condition on the processes through which brown dwarfs can form. These formation processes may be similar to mechanisms that drive star formation like gravitational collapse of giant molecular clouds, or analogous to those of giant planets through planetesimal growth in a circumstellar disk. Brown dwarfs could also form through entirely other modes such as ejection from star forming aggregates \citep{RC01}. 

Studies of the multiplicity of these objects is an efficient approach to constrain theories of their formation. Those studies can measure and constrain fundamental trends such as the orbital separation of binaries, $a$, as well as the companion-to-host mass ratio, $q \equiv M_{\rm B}/M_{\rm A}$. For stars the orbital separation distribution appears log-normal with a larger peak for higher mass stars \citep[e.g.][]{Winters+2019, Raghavan+2010, DeRosa+2014}, so that the binary separation increases with mass. The companion mass ratio distribution also appears to change as a function of primary mass \citep{DK13}, changing from bottom heavy to top heavy for O-stars to brown dwarfs respectively \citep{Offner+2022}. Brown dwarf binaries appear to be consistent with the orbital radius trend observed for higher mass multiples and typically form tighter orbital configurations. \citet{Burgasser+2003, Burgasser+2006} studied T0 to T8 spectral types for binarity, finding all identified systems to have separations $<$5 au and mass ratios $>0.8$, thus confirming the trend. Those early studies were however limited by only being sensitive to mass ratios $>0.5$ and did not probe the full range of mass ratios. More recent works have been able to complement this, confirming the trend over a larger range of $q = 0.2$-1 and separations $a = 0.1$-1000 au \citep{Fontanive+2018}, hence extending previous works across virtually the full dynamic range of mass ratios and separations.

During the last decade we have seen an increase in the discoveries of brown dwarfs in the solar neighborhood, much to the efforts of deep imaging surveys and infrared missions such as the Wide-field Infrared Survey Explorer \citep[WISE,][]{WISE}. \citet{Cushing+2011} discovered several brown dwarfs from WISE data to have spectral features which distinguished them from the previously known latest T-type dwarfs, providing a clear transition to the identified Y dwarf spectral class. These Y dwarfs represent an extreme in temperature among the field star populations of the Milky Way \citep{Cushing+2011, Kirkpatrick+2019, Zalesky+2019}. They have effective temperatures lower than 500\,K, with some approaching that of Jupiter \citep[130 K,][]{Hanel+1981}. For brown dwarfs, Coulomb and electron degeneracy effects compete in the equation of state, dictating their structure so that typical radii are close to that of Jupiter \citep{Burrows+2011}. Thus, the range of temperatures results in a range of luminosities.

Y dwarfs are much brighter at wavelengths $>4.5\,\mu$m than at shorter wavelengths, so for these objects the Near Infrared Camera (NIRCam) on JWST is orders of magnitude more sensitive than any other facility and the ideal tool for studying Y-dwarf multiplicity properties. As part of a JWST Cycle 1 GO program \citep[\#2473,][]{Albert+2021}\footnote{\url{https://www.stsci.edu/jwst/phase2-public/2473.pdf}}, we are conducting a survey of 20 Y dwarfs that is sensitive to companions beyond 1~au and down to $1\,M_{\rm Jup}$ given estimated sensitivity limits pre-launch \citep{Ceau+2019}. Here we present our first discovery, a faint companion to the Y dwarf WISE J033605.05$-$014350.4 (hereafter W0336) and therefore, the first Y+Y binary system.

\section{Observations and reduction}\label{sec:observations}
\noindent W0336 was first discovered by \citet{Kirkpatrick+2012} using WISE data and the unresolved system has been classified as a Y0 dwarf \citep{Martin+2018}\footnote{Some discrepancies have been reported in the W0336 spectral type, likely related to a low signal to noise ratio of the observed spectrum \citep{Mace+2013, Martin+2018}.} with an estimated effective temperature of $T_{\rm eff} = 460 \pm 79$~K \citep{Kirkpatrick+2021}. Spitzer \citep{Spitzer} observations were used to measure the astrometry of the object \citep{Martin+2018, Kirkpatrick+2019}, and here we adopt the results from \citet{Kirkpatrick+2021} with a parallax of $\pi = 99.8 \pm 2.1$ mas and proper motions of $\mu_{\rm RA} = -251.5 \pm 0.9$ mas/yr and $\mu_{\rm DEC} = -1216.1 \pm 0.93$ mas/yr.

We observed the target with JWST/NIRCam on 2022 September 22 at 12:49:44 UT, obtaining images in both F150W and F480M filters simultaneously. The source was observed with NIRCam imaging mode using a 5-point subpixel Small-Grid-Dither pattern for a total exposure time of 2630.509 seconds with the BRIGHT1 readout pattern. The data were reduced by the official JWST pipeline with the data processing software version 2022\_3a, and downloaded from the MAST archive based on calibration software version 1.7.2\footnote{Using calibration reference data system version {\it 11.16.14} with the reference file {\it jwst\_1023.pmap}.}. The unresolved photometry of the system was estimated using the photoutils Python package \citep{Bradley+2022} and the JWST photometric reference data {\it jwst\_nircam\_photom\_0114.fits}. An aperture size equal to $70\,\%$ encircled energy was adopted together with appropriate aperture corrections in accordance to the JWST Calibration Reference Data System. We employed our analysis on the intermediate data reduction step, level 2 pipeline product {\it *cal.fits} files, further described in Section~\ref{sec:ePSF}. In addition to the standard calibrated images we performed our own background subtraction by fitting an annulus with 10--15 pixel radius centered upon the target.

\begin{figure}
    \centering
    \includegraphics[width=0.99\linewidth]{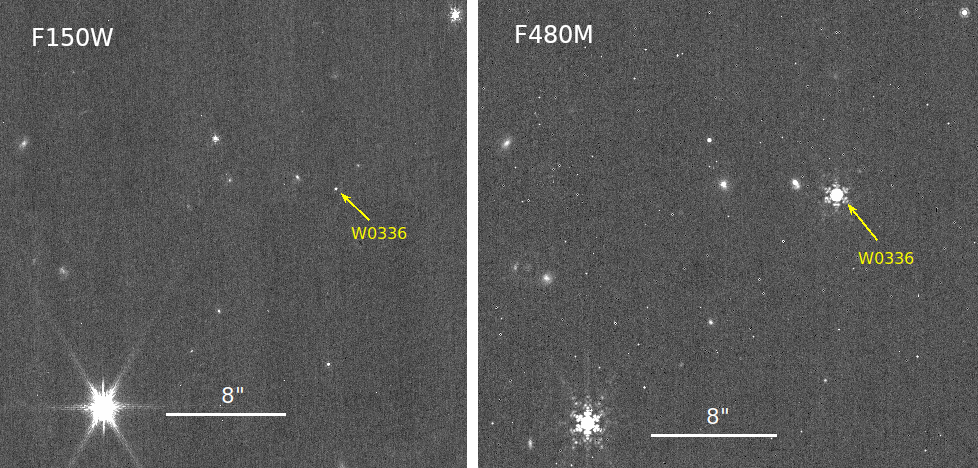}
    \caption{JWST/NIRCam images showing the aligned and scaled images of the target W0336 in each observed filter. North is up and East is to the left in the images.}
    \label{fig:target}
\end{figure}

We verified the positions of the target W0336 for the two bands by overplotting their respective images. The images from each band are shown in Figure~\ref{fig:target}, illustrating that W0336 appears brighter in the longer wavelength band F480M. The images have been rotated with respect to the instrument aperture positional angle (NIRCam APA) of $271.61^{\circ}$ in the F150W band and $270.79^{\circ}$ in the F480M band at the time of the observations, so that North is up and East is to the left in the images. We also cross-referenced the position of W0336 with publicly available Hubble Space Telescope images from GO-16229\footnote{\url{https://www.stsci.edu/hst/phase2-public/16229.pdf}} in the F160W filter, verifying the position and motion of the system compared to background sources.

%%%%%%%%%%%%%%%%%%%%
%%%%%%%%%%%%%%%%%%%%

\section{Empirical point spread function analysis}\label{sec:ePSF}
\noindent In order to search for companions at small separations, we applied a binary-PSF fitting routine using empirically derived PSF models, thoroughly described in \citet{DeFurio+2023}. This analysis was based on the empirical PSF (ePSF) construction described by \citet{AK00} and has been applied to find close companions on sub-pixel scales using for example Hubble Space Telescope images \citet{DeFurio+2019, DeFurio+2022a, DeFurio+2022b}.

We constructed the ePSF for each filter using three other Y-dwarfs which were observed close in time to W0336 (WISE J035934.06$-$540154.6, WISE J030449.03$-$270508.3, and WISE J041022.71$+$150248.4), within a 36 hour window. We used a pixel box size of $11\times11$ pixels centered upon the flux peak of the selected sources, and constructed a $4\times$ oversampled PSF model. Each source was observed with a pattern of five sub-pixel dither positions, allowing for more detailed sub-pixel modeling of the ePSF. All dither positions were used in making of the ePSF for a total of $3\times5=15$ sources for the final combined ePSF in each filter. The PSF for a relatively wide bandpass can vary between objects if those objects have different spectral energy distributions through the filter. We therefore chose these sources to ensure consistent color to retain the integrity of the expected PSF of a Y-dwarf and to minimize wavefront error drifts over a short amount of time in order to optimize our estimation of the astrophysical scene.

We then used the ePSF to fit both single and double PSF models to the Y-dwarf data. We employed a Nested Sampling routine using the PyMultiNest Python package \citep{Buchner+2014} that performs the Nested Sampling Monte Carlo analysis using MultiNest \citep{Feroz+2009}. Our single model consists of three parameters: $x$ and $y$ center position and flux normalization of the target source. Our binary model consists of six parameters: $x$ and $y$ center of the primary, flux normalization of the primary, separation between the centers of the primary and secondary, the position angle of the center of the secondary relative to the primary, and the flux ratio between the secondary and primary $(f_{\rm r} \equiv f_{\rm B}/f_{\rm A})$. The priors to the primary coordinates were set as flat over $\pm 1.5$ pixels from the centre of the pixel box. The companion parameter priors were set as flat over separations between 0.5-5.0 pixels, as we do not expect to be sensitive to companions closer in than half a pixel and we limited the search to the size of the pixel box. The positional angle prior was set to be flat between 0-360$^{\circ}$ and the flux ratio prior log-uniform over $10^{-4}$-1. 

\begin{figure}
    \centering
    \includegraphics[width=0.99\linewidth]{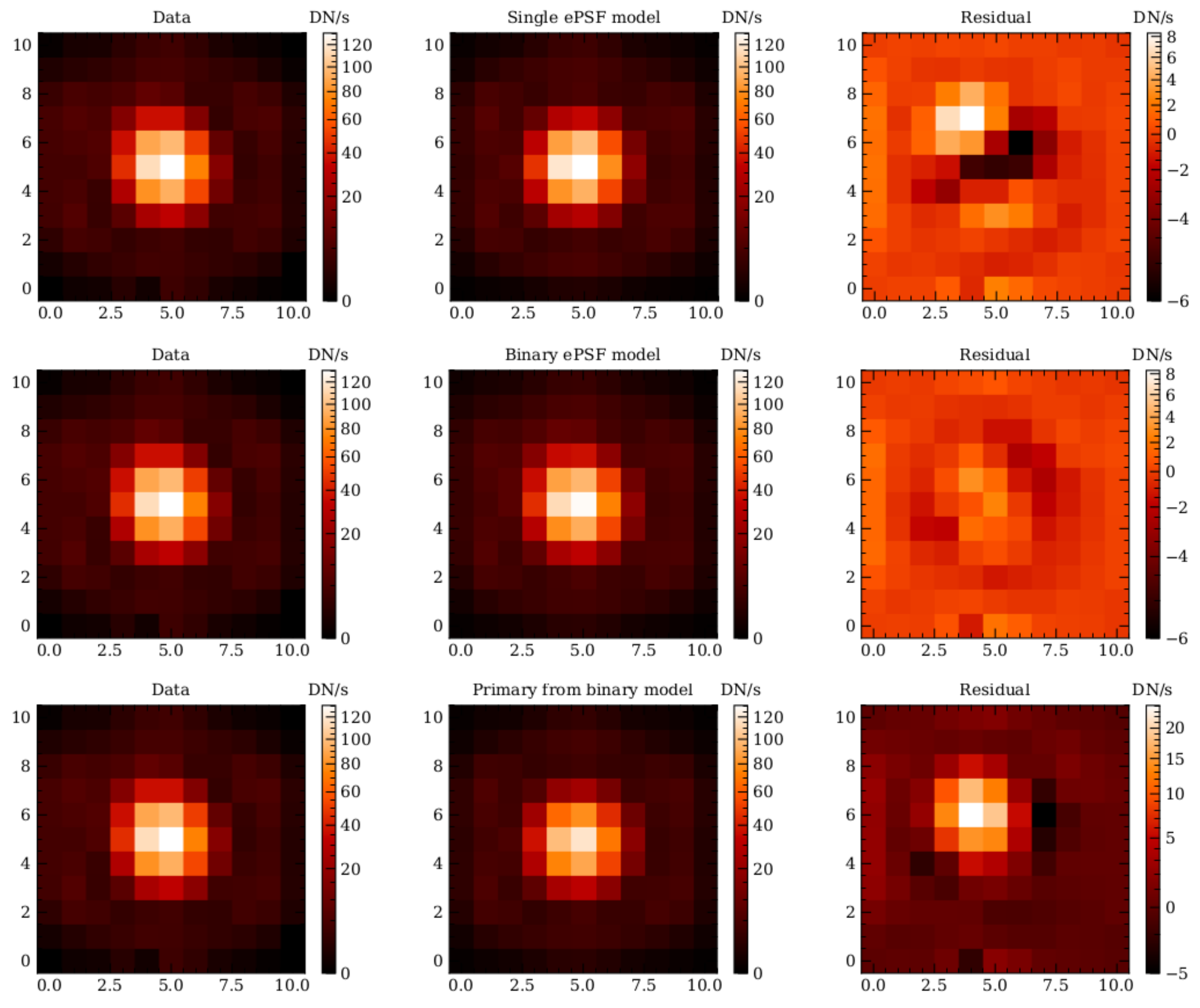}\\
    \caption{Images displaying the pipeline calibrated data of W0336 in the F480M band in the left column, models in the middle column, and their corresponding residuals when the models have been subtracted from the data in the right column. The top row shows a single model fitted to the data and the middle row a binary double ePSF model. The bottom row depicts the same binary model as the middle row, but only showing the primary component from that fit to better highlight the companion seen in the residuals after subtracting the primary component from the data. The units are in DN/s. The color scheme in the images are scaled to a powerlaw with exponent of 0.5, and the colorbar for the binary model residual image has been scaled to match the single model residual image to better highlight the smaller residual and improved fit.}
    \label{fig:datmodres_F480M}
\end{figure}

PyMultiNest calculates the Bayesian evidence of the model, which we used to determine whether a binary or single model fits the data better. We first fit each dither position individually to ensure that the companion was detected in every image, while also verifying consistency and check for possible bad frames. We then fit all dither positions simultaneously, treating the $x$- and $y$- positions of the primary as free parameters for each dither position while keeping the parameters for the secondary component the same for all dithers, for a total of 14 parameters to be varied in the binary model fit. Figure~\ref{fig:datmodres_F480M} displays the data images, model systems and corresponding residuals for the different cases in the F480M filter, and the best fit values for both filters are presented in Table~\ref{tab:results}.

%%%%%%%%%%%%%%%%%%%%
%%%%%%%%%%%%%%%%%%%%

\section{Results and discussion}\label{sec:results}
\noindent The empirical PSF analysis provides a compelling first look at these data. This is the only companion found with this method in our sample of 12 Y dwarfs observed to date, and the first companion ever discovered around a Y dwarf primary. Although these are low number statistics, the observed frequency of companions in our sample (1/12) is consistent with the companion frequency to late T-dwarfs (8\%) over mass ratios of $q=$0.2-1.0 and separations of $a=$0.1-1000 au \citep[e.g.][]{Fontanive+2018}. Future work will characterize the companion population and estimate the companion frequency over mass ratios and orbital separations sampled across our entire survey.

%%%%%%%%%%%%%%%%%%%%
\begin{figure}
    \centering
    \includegraphics[width=0.6\linewidth]{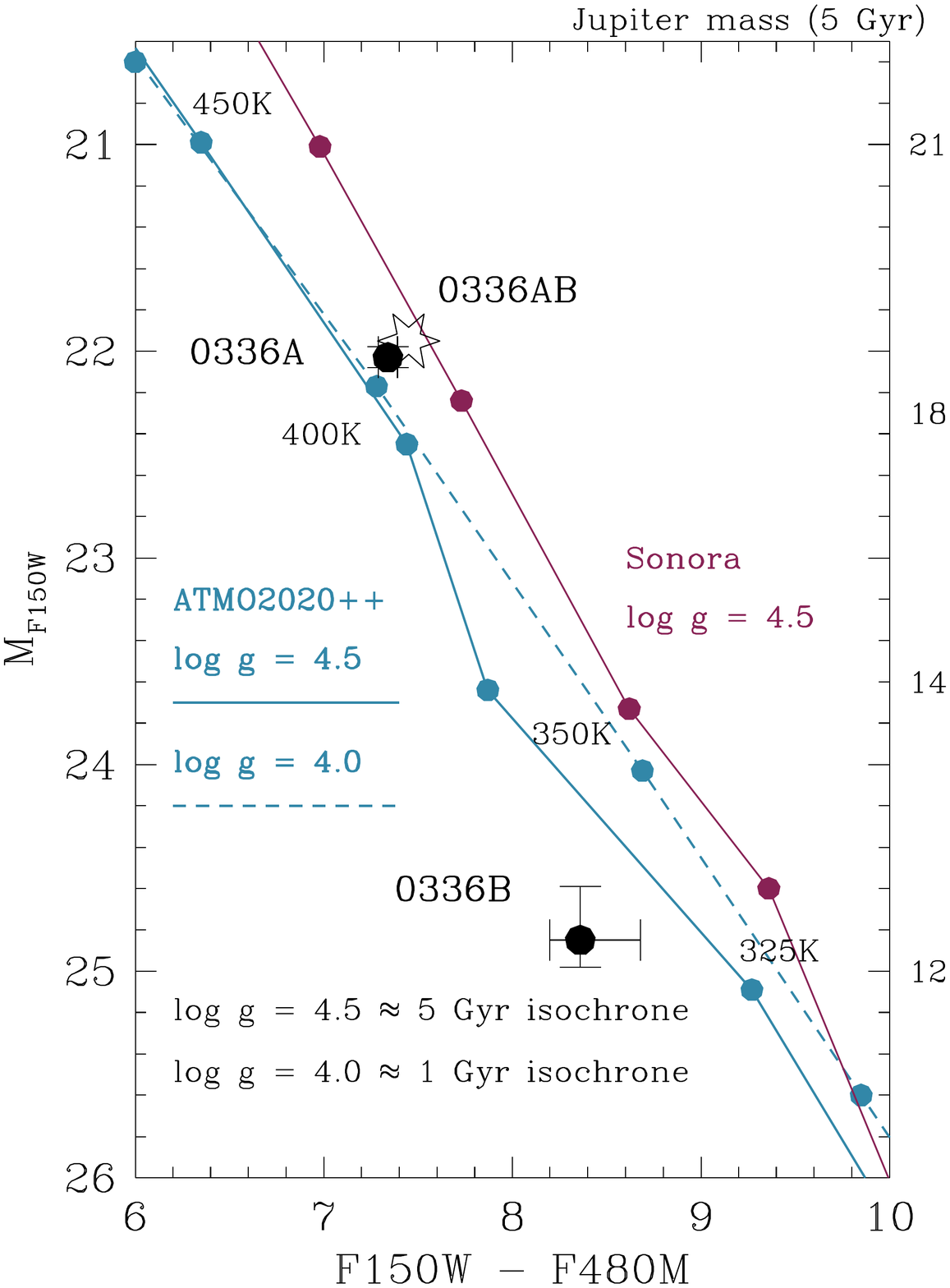}
    \caption{Color-magnitude diagram for W0336 in the NIRCam filters used in this work. The open star corresponds to the unresolved system, and black filled points indicate the resolved components. Model sequences are shown for effective temperatures and surface gravities indicated in the legend. All models are for cloud-free solar metallicity atmospheres, however the Sonora models \citep{Marley+2021} are in chemical equilibrium, while the ATMO2020++ models \citep{Phillips+2020, Leggett+2021} include disequilibrium chemistry and also an adjusted pressure-temperature profile. The masses along the right axis are taken from the Sonora evolutionary models of \citet{Marley+2021}, for log g $=$ 4.5 and the temperatures indicated along the sequences. }
    \label{fig:CMD}
\end{figure}
%%%%%%%%%%%%%%%%%%%%

Table~\ref{tab:results} gives the angular separation and orientation of the resolved system. The values are estimated from the median of the probability density function obtained from the multinested Monte Carlo approach, with errors being the $16\,\%$ and $84\,\%$ percentiles. The separation measured in pixels is converted into milliarcseconds using the nominal plate scale from the JWST user documentation\footnote{\url{https://jwst-docs.stsci.edu/jwst-near-infrared-camera}} of 63 mas/pixel and 31 mas/pixel for the F480M and F150W filters respectively. We adopt the angular separation of $0{\farcs}084$ measured in the F480M filter where the companion is brighter. Given the distance of $10.02\pm0.21$~pc from the Spitzer parallax, the estimated separation corresponds to a projected separation of $0.89^{+0.07}_{-0.10}$~au. This separation is consistent with what one would expect given the orbital separation distribution from \citet{Fontanive+2018} for late T and early Y dwarfs with a peak at $2.9^{+0.8}_{-1.4}$ au in projected separation. The probability of the companion being a background contamination is extremely low given its proximity to the primary. Assuming a randomly uniform distribution of objects brighter than the companion W0336B in the field of view of our NIRCam images, we obtain a probability of $\sim 4\times10^{-6}$ for a chance projection at the same separation as the discovered companion.

Table~\ref{tab:results} also gives the measured F150W and F480M Vega magnitudes for the unresolved system and for the resolved components, where the resolved magnitudes are calculated from our contrast measurements of flux ratio between the components. Figure~\ref{fig:CMD} shows a color-magnitude diagram for the system, together with model sequences. There are two recent model families which apply cloud-free atmospheres and cover the low luminosities of the W0336 system - the Sonora Bobcat models \citep{Marley+2021} and the ATMO2020 models \citep{Phillips+2020, Leggett+2021}. The ATMO2020 models include disequilibrium chemistry and a physically-motivated empirical adjustment to the atmosphere's pressure-temperature profile \citep[see discussion in][]{Leggett+2021}. Figure~\ref{fig:CMD} shows that the two model sets predict very similar values of $M_{\rm F150W}$ for a given $T_{\rm eff}$ and gravity, although the F150W $-$ F480M colors can differ. The models are thus expected to provide reliable estimates of $T_{\rm eff}$ for the W0336 system, given in Table~\ref{tab:results}. The uncertainties in $T_{\rm eff}$ have been estimated from the range of $M_{\rm F150W}$ values calculated by the models for plausible ranges in gravity and metallicity \citep[Table 9 of ][]{Leggett+2021}.

There are no constraints on age for this isolated system. However studies of the solar neighborhood, and of low mass stars and brown dwarfs in particular, suggest a likely age range for this system of 1-3~Gyr \citep[e.g.][]{DL17}. The ATMO2020 models and Figure~\ref{fig:CMD} on the other hand suggest that the age of the system may be older than 5~Gyr, and also that the system has approximately solar metallicity (not shown). However, the uncertainty in the measured F150W $-$ F480M colors, and systematic differences in the models, means that surface gravity cannot be fully constrained, which would have in turn constrained mass and hence age.

Nevertheless, Table~\ref{tab:results} gives estimates of mass for the W0336 components for ages of 1~Gyr and 5~Gyr, derived from our absolute magnitude estimates and the Sonora evolutionary models \citep{Marley+2021}. The mass-estimates range from 7.5-$20.0\,M_{\rm Jup}$ for the primary and 4.0-$12.5\,M_{\rm Jup}$ for the secondary, and the mass ratio of the system is around $q=0.6$, at the low end of the distribution found for low-mass binaries with masses $\leq 40\,M_{\rm Jup}$ inferred from Lyon/COND models \citep{Fontanive+2018}. Interestingly, W0336B appears to lie at or below the deuterium burning limit of $13\,M_{\rm Jup}$, sometimes used as a boundary between brown dwarfs and planets \citep{Spiegel+2011}. If the system is 2~Gyr old or younger, both components lie below this limit. In any case, the companion joins a growing list of cold brown dwarfs and isolated young brown dwarfs which may serve as a constraint on the minimum mass for opacity limited fragmentation.

Applying the conversion factor of 1.16 between projected and physical orbits from \citet{DL11} for isotropic randomly distributed inclinations with the measured distance, we find the W0336 binary to be separated by $0.97^{+0.05}_{-0.09}$ au. Combining this with the mass-range from the evolutionary models, a tentative orbital period can be estimated for the system of 5-6 years in the higher system mass scenario, and 7-9 years in the lower system mass case. A dynamical mass from orbital monitoring could thus be obtained in a relative short time span for the system. The system is far too faint for Gaia to observe any astrometric acceleration that could otherwise aid to place dynamical mass constraints.

Among the few brown dwarf binaries where the primary has spectral type of T8 or later, few have been reported to be low $q$ systems: WISEPC J121756.91$+$162640.2 (T9$+$Y0, 12$+$8 $M_{\rm Jup}$) and WISEPA J171104.60$+$350036.8 \citep[T8$+$T9.5, 20$+$9 $M_{\rm Jup}$][]{Liu+2012}. However, both these systems have much larger separations of $\sim 8$-15 au and break from the expected orbital distribution trend otherwise observed. \citet{Dupuy+2015} reported on the T+Y dwarf binary WISE J014656.66$+$423410.0, with a similar projected separation as W0336 of $0.93^{+0.16}_{-0.12}$~au, albeit with a higher mass ratio $q\geq0.9$. That W0336 in contrast is a tightly bound system with low mass-ratio is an intriguing discovery. With the low number statistics for these extreme systems it is unclear whether they represent the true binary population of ultracool dwarfs or can be considered as peculiar systems. It is also possible that many systems go undetected where the separation is too tight for them to be retrieved. Indeed, W0336 confirms that late-type and low-mass binaries exist and survive in tight configurations, and surveys approaching sub-au separations are needed to discover them. Such endeavors could be taken using for example Kernel Phase Interferometry \citep{Ceau+2019}.

Follow-up spectroscopic and photometric observations over a wide wavelength range are needed to further constrain the properties of both the primary and its companion. Substellar binaries on small separations such as these can be resolved with for example integral field units to obtain valuable spectral information \citep[e.g.][]{Calissendorff+2019}. Such multiple pairs, assuming a common formation age and composition, represent critical constraints on atmospheric as well as evolutionary theory of the formation and evolution of substellar companions.

%%%%%%%%%%%%%%%%%%%%
\begin{table*}[t]
\renewcommand{\arraystretch}{1.25}
\centering
\caption{Properties of the W0336 binary system}
\begin{tabular}{lcc}
\hline \hline
Band & F150W & F480M \\
Separation [mas] & $89.8^{+3.8}_{-4.1}$ & $83.7^{+4.9}_{-8.2}$ \\
Position angle [deg] & $299.1 \pm 3.4$ & $295.4^{+2.3}_{-2.6}$ \\
Contrast [mag] & $2.82^{+0.19}_{-0.11}$ & $1.81^{+0.14}_{-0.31}$ \\
%\hline
%Component & Primary & Secondary \\
W0336 AB & $21.97 \pm 0.01$ & $14.52 \pm 0.01$ \\
W0336 A & $22.05 \pm 0.01$ & $14.71^{+0.02}_{-0.05}$ \\
W0336 B & $24.87^{+0.18}_{-0.10}$ & $16.51^{+0.12}_{-0.26}$ \\
\hline
Component & Primary & Secondary \\
$T_{\rm eff}$ [K]& $415 \pm 20$ & $325^{+15}_{-10}$ \\
$M$ [$M_{\rm Jup}$] (1 Gyr) & $8.5\pm1$ & $5\pm1$ \\
$M$ [$M_{\rm Jup}$] (5 Gyr) & $18\pm2$ & $11.5\pm1$ \\
\hline
\multicolumn{1}{l}{Physical separation [au]} & \multicolumn{2}{c}{$0.97^{+0.05}_{-0.09}$}  \\
\multicolumn{1}{l}{Orbital period [years]} & \multicolumn{2}{c}{$7 \pm 2$ }  \\
\multicolumn{1}{l}{Mass fraction $q=M_{\rm B}/M_{\rm A}$} & \multicolumn{2}{c}{$0.61 \pm 0.05$}  \\
\hline
\label{tab:results}
\end{tabular}
\end{table*}
%%%%%%%%%%%%%%%%%%%%

\section*{Acknowledgements}
\noindent We thank the reviewer for the constructive comments. This work is based on observations made with the NASA/ESA/CSA James Webb Space Telescope. The data were obtained from the Mikulski Archive for Space Telescopes at the Space Telescope Science Institute, which is operated by the Association of Universities for Research in Astronomy, Inc., under NASA contract NAS 5-03127 for JWST. These observations are associated with program \#2473 and can be accessed via \url{10.17909/9zgc-m183}. L.A. acknowledges support of the Canadian Space Agency through contract 22JWGO1-11. T.L.R. would like to acknowledge the support of the NASA Science Mission Directorate under WBS 411672.07.05.05.03.01. This work was authored by employees of Caltech/IPAC under Contract No. 80GSFC21R0032 with the National Aeronautics and Space Administration.

\end{document}